\documentclass{emulateapj}

\usepackage{epsfig}
\usepackage{longtable}

\begin{document}

\title{Spectroscopic Confirmation of Two Massive
  Red-Sequence-Selected Galaxy Clusters at \lowercase{$z$} $\sim$ 1.2 in the
  SpARCS-North Cluster Survey}

\author{Adam Muzzin\altaffilmark{1}, Gillian Wilson\altaffilmark{2},
  H.K.C. Yee\altaffilmark{3}, Henk Hoekstra\altaffilmark{4,5,6},
  David Gilbank\altaffilmark{7}, Jason
  Surace\altaffilmark{8}, Mark Lacy\altaffilmark{8}, Kris
  Blindert\altaffilmark{9}, Subhabrata Majumdar\altaffilmark{10},
  Ricardo Demarco\altaffilmark{2}, Jonathan P. Gardner\altaffilmark{11},
  Mike Gladders\altaffilmark{12} \& Carol Lonsdale\altaffilmark{13}}

\altaffiltext{1}{Department of Astronomy, Yale
  University, New Haven, CT, 06520-8101; adam.muzzin@yale.edu} 
\altaffiltext{2}{Department of Physics and Astronomy,
University of California, Riverside, CA 92521}
\altaffiltext{3}{Department. of Astronomy \& Astrophysics, University
  of Toronto, 50 St. George St., Toronto, Ontario, Canada, M5S 3H4} 
\altaffiltext{4}{Department of Physics and Astronomy, University of
  Victoria, Victoria, BC V8P 5C2, Canada}
\altaffiltext{5}{Alfred P. Sloan Fellow}
\altaffiltext{6}{Leiden Observatory, Leiden University, PO Box 9513,
  2300RA Leiden, The Netherlands}
\altaffiltext{7}{Astrophysics and Gravitation Group, Department of
  Physics \& Astronomy, University of Waterloo, Waterloo, Ontario,
  Canada N2L 3G1} 
\altaffiltext{8}{Spitzer Science Center, California Institute
  of Technology, 220-6, Pasadena, CA, 91125} 
\altaffiltext{9}{Max Planck Institute for Astronomy
Koenigstuhl 17, 69117, Heidelberg, Germany} 
\altaffiltext{10}{Department of Astronomy and Astrophysics, Tata Institute of Fundamental Research 
1, Homi Bhabha Road, Colaba, Mumbai 400 005, India}
\altaffiltext{11}{Goddard Space Flight Center, Code 665,
Laboratory for Observational Cosmology, Greenbelt MD 20771}
\altaffiltext{12}{University of Chicago, 5640 South Ellis Avenue,
  Chicago, IL 60637}
\altaffiltext{13}{North American ALMA Science Center, NRAO Headquarters, 520 Edgemont Road, Charlottesville, VA 22903}
\begin{abstract}
The Spitzer Adaptation of the Red-sequence Cluster Survey (SpARCS) is
a deep z$^{\prime}$-band imaging survey covering the $Spitzer$ SWIRE
Legacy fields designed to create the first large homogeneously-selected sample of massive clusters at $z >$ 1 using an
infrared adaptation of the cluster red-sequence method.   We present an
overview of the northern component of the survey which has been
observed with CFHT/MegaCam and covers
28.3 deg$^2$.  The southern component of the survey was observed with
CTIO/MOSAICII, covers 13.6 deg$^2$, and is summarized in a companion paper by Wilson et
al. (2008).  We also present spectroscopic confirmation of two rich
cluster candidates at $z \sim 1.2$.  
Based on Nod-and-Shuffle spectroscopy from  GMOS-N on Gemini there are
17 and 28 confirmed cluster members in SpARCS J163435+402151 and
SpARCS J163852+403843 which have spectroscopic redshifts of 1.1798 
 and 1.1963, respectively.  The clusters have velocity dispersions of
490 $\pm$ 140 km/s and 650 $\pm$ 160 km/s, respectively which imply masses (M$_{200}$) of (1.0 $\pm$ 0.9) x
10$^{14}$ M$_{\odot}$ and (2.4 $\pm$ 1.8) x
10$^{14}$ M$_{\odot}$.  Confirmation of these candidates
as $bona fide$ massive clusters demonstrates that two-filter imaging
is an effective, yet observationally efficient, method for selecting clusters at $z >$ 1.
\end{abstract}


\section{Introduction}
In the nearby universe there are numerous lines of evidence  suggesting that environmental
processes could be the dominant force driving the evolution of
the galaxy population.  Properties such as star formation rate (SFR, e.g.,
Lewis et al. 2002; Gomez et al. 2003; Kauffmann et al. 2004),
morphology (e.g., Dressler 1980; Goto et al. 2003; Park et al. 2007),
stellar mass (e.g., Kauffmann et al. 2004),
color (e.g., Hogg et al. 2003; Balogh et
al. 2004;  Blanton et al. 2005), and luminosity (e.g., Croton et al. 2005; Park
et al. 2007) are all strongly correlated  with local galaxy density.   Although there is still
 debate about which, if any, of these relations are 
``fundamental''  (e.g., Hogg et al. 2004; Park et al. 2007),
it is clear that the mean properties of galaxies we measure depend
strongly on the type of  environment they occupy.
\newline\indent
An obvious first step toward a better understanding of environmental
processes is to study  how they
evolve with redshift.  At higher redshift, the overall population of
galaxies is younger and have been living within their local
environment for less time.  The environmental processes that are most
effective and have the shortest timescales should be  most apparent
when comparing galaxies at different densities in the high redshift
universe.  The data at higher redshift are still somewhat sparse
compared to the nearby universe but it is beginning to emerge that
properties such as the SFR (e.g., Elbaz et al. 2007; Cooper et al. 2008;
Poggianti et al. 2008), color (e.g., Cooper et al. 2007) and morphology
(e.g., Dressler et al. 1997; Postman et al. 2005; Smith et al. 2005,
Capak et al. 2007) are still correlated with local density, albeit differently from the nearby universe.
\newline\indent
Of particular interest for understanding environmental processes are the
cores of rich galaxy clusters.  These are the most extreme density environments at
all redshifts, and if environment is truly an important force in galaxy
evolution a comparison of the properties of galaxies that live in this
environment to those that live in the field should provide the largest
contrasts.  
Despite their potential value for such
studies, and the abundance of resources directed at finding distant clusters,
there are still relatively few confirmed rich clusters at $z >$ 1.
\newline\indent
The major challenge for cluster surveys targeting the $z >$ 1 range is the need to be
simultaneously deep enough to detect either the galaxies or hot X-ray gas in clusters
and yet  wide enough to  be able to cover a large area because 
of the rarity of rich clusters at $z >$ 1.  
The requirement of both depth and area  has pushed X-ray detection of
clusters with current telescopes to the limit.  
\newline\indent
The largest area targeted X-ray cluster surveys are
the XMM-LSS (Valtchanov et al. 2004; Andreon et al. 2005; Pierre et al. 2006) and the XMM-COSMOS (Finoguenov et
al. 2007), and while these have been successful at discovering
$z >$ 1 clusters (e.g., Bremer et al. 2006), they cover areas of only 9 and 2 deg$^2$,
respectively and are therefore limited to fairly low mass systems on
average.  Indeed, X-ray detection of clusters at $z >$ 1 is so challenging
that currently the most promising surveys are those searching for
clusters serendipitously in the $entire$ XMM-Newton archive (e.g., Romer et al. 2001; Mullis et
al. 2005; Stanford et al. 2006; Lamer et al. 2008).  
\newline\indent
Complementary to X-ray detection is optical detection of clusters using
overdensities of galaxies selected using  the
red-sequence (e.g., Gladders \& Yee 2000, 2005; Gilbank et al. 2004; Muzzin et al. 2008) or
photometric redshifts (e.g., Stanford et al. 2005; van Breuklen et
al. 2007; Eisenhardt et al. 2008).  Recently, it has become clear that the key to discovering clusters
above $z >$ 1 with these techniques is the incorporation of
infrared (IR) data which probes the peak of the stellar emission for
galaxies at $z >$ 1.
\newline\indent
Although  IR surveys have thus far been confined to 
modest areas (ranging from 0.5 - 8.5 deg$^2$) they have been extremely
successful at detecting $z >$ 1 clusters (e.g., Stanford et al. 2005;
Brodwin et al. 2006; van Breuklen et al. 2007; Zatloukal et al. 2007;
Krick et al. 2008; Eisenhardt et al. 2008; Muzzin et al. 2008).  
The IR cluster community is now regularly discovering clusters at $z >$ 1
and shortly IR-detected clusters  should
outnumber their X-ray counterparts.
\newline\indent
Currently the largest area IR survey still deep enough to detect clusters at $z
>$ 1 is the $Spitzer$ Wide-Area Infrared
Extragalactic Survey (SWIRE, Lonsdale et al. 2003; Surace et al. 2005).  SWIRE covers
$\sim$ 50 deg$^2$ in the $Spitzer$ bandpasses and is slightly deeper,
and nearly a factor of six larger than the next largest IR cluster survey, the
IRAC Shallow Survey Cluster Search (ISCS, Eisenhardt et al. 2008).  
\newline\indent
In Muzzin et al. (2008) we demonstrated the potential of using the
red-sequence method with $Spitzer$ data to detect distant clusters
using data from the 3.8 deg$^2$ $Spitzer$ First Look Survey (FLS, Lacy et
al. 2004).
In 2006 we began observations for the $Spitzer$ Adaptation of the
Red-sequence Cluster Survey (SpARCS), a deep
z$^{\prime}$-band imaging survey of the SWIRE fields. SpARCS aims
to discover the first large, yet homogeneously-selected sample of rich
clusters at $z >$1 using the red-sequence method.  SpARCS is similar
to the RCS surveys (Gladders \& Yee 2005; Yee et al. 2007) which target clusters to $z \sim$
1 using an R - z$^{\prime}$ color except that we use a z$^{\prime}$ -
3.6$\micron$ color, which spans the 4000\AA~break at $z >$ 1.
With a total area effective area of 41.9 deg$^2$ SpARCS is currently the only $z
>$ 1 cluster survey that can discover a significant number of rare rich clusters.  These clusters will be extremely
valuable for quantifying the evolution of galaxy
properties in the densest environments at high redshift.
\newline\indent
This paper is organized as follows.  In $\S$ 2 we provide a brief
overview of the northern component of the SpARCS survey (the southern
component is summarized in Wilson et al. 2008).  In $\S$ 3 we discuss
the selection of cluster candidates that were chosen for followup
spectroscopy, and in $\S$ 4 we present spectroscopic confirmation of
two $z >$ 1 clusters from early SpARCS data.  In $\S$ 5 we present the
dynamical analysis of the clusters followed by a discussion of the
cluster properties in $\S$ 6.  We conclude with a summary in $\S$ 7.
\newline\indent
Throughout this paper we assume an $\Omega_{m}$ = 0.3,
$\Omega_{\Lambda}$ = 0.7, H$_{0}$ = 70 km s$^{-1}$ Mpc$^{-1}$
cosmology.  All magnitudes are on the Vega system unless indicated
otherwise.
\section{The SpARCS-North Survey}
The SWIRE survey is located in six fields and contains $\sim$ 50 deg$^2$ of imaging in the
  four IRAC bandpasses (3.6$\micron$, 4.5$\micron$, 5.8$\micron$, and
  8.0$\micron$) and the three MIPS bandpasses (24$\micron$,
  70$\micron$, and 160$\micron$).  Three of the fields are located in
  the northern hemisphere (ELAIS-N1, ELAIS-N2, and the Lockman Hole),
  two of the fields are located in the southern hemisphere (ELAIS-S1,
  and the Chandra-S), and one of the fields is equatorial, the XMM-LSS
  field.  A thorough discussion of the data reduction,
photometry, cluster finding, and the SpARCS catalogue for all fields will be
presented in a future paper by A. Muzzin et al. (2009, in
preparation).  Here we present a brief summary of the z$^{\prime}$-band observations of the ELAIS-N1,
  ELAIS-N2, Lockman Hole, and XMM-LSS\footnote{The XMM-LSS data was
  obtained as part of the CFHT Legacy Survey} fields obtained with CFHT/MegaCam; hereafter the
  SpARCS-North Survey.  Observations of the ELAIS-S1 and Chandra-S fields
  were obtained with the CTIO/MOSAICII and are outlined in the
  companion paper by Wilson et al. (2008). 
\newline\indent
The IRAC imaging of the ELAIS-N1, ELAIS-N2, Lockman, and XMM-LSS
fields covers areas of 9.8, 4.5, 11.6, and 9.4 deg$^2$, respectively.  In
Figure 1 we plot the  IRAC 3.6$\micron$ mosaics for
these fields.  The superposed white squares  
represent the locations of the CFHT/MegaCam pointings.  The pointings
were designed to maximize the overlap with the IRAC data, but to minimize
the overall number of pointings by omitting regions that have little overlap with the IRAC
data.  
There are a total of 12, 5, 15, and 13 MegaCam pointings in the ELAIS-N1,
ELAIS-N2, Lockman Hole and XMM-LSS fields, respectively.
\newline\indent
We obtained observations in the z$^{\prime}$-band with 6000s of integration time for each pointing in the ELAIS-N1,
ELAIS-N2 and Lockman Hole fields in queue mode using CFHT/MegaCam which is composed of
36 4096 $\times$ 2048 pixel CCDs, and has a field of view (FOV) of
$\sim$ 1 deg$^2$.
Omitting the large chip gap areas and regions contaminated by bright stars, the total overlap
region with both z$^{\prime}$ and IRAC data is 28.3 deg$^2$.
\newline\indent
Photometry was performed on both the z$^{\prime}$ and IRAC mosaics
using the SExtractor photometry package (Bertin \& Arnouts 1996).  Colors were determined using 3 IRAC pixel (3.66$''$)
diameter apertures.  The IRAC data was corrected for flux lost outside
this aperture due to the wings of the PSF using aperture corrections
measured by Lacy et al. (2005).  Total magnitudes for the IRAC photometry were computed using
the method outlined in Lacy et al. (2005) and Muzzin et al. (2008).
The 5$\sigma$ depth of the z$^{\prime}$ data varies depending on the seeing and
the sky background; however, the mean 5$\sigma$ depth for extended sources is z$^{\prime}$
$\sim$ 23.7 Vega (24.2 AB).
\begin{figure*}
\plotone{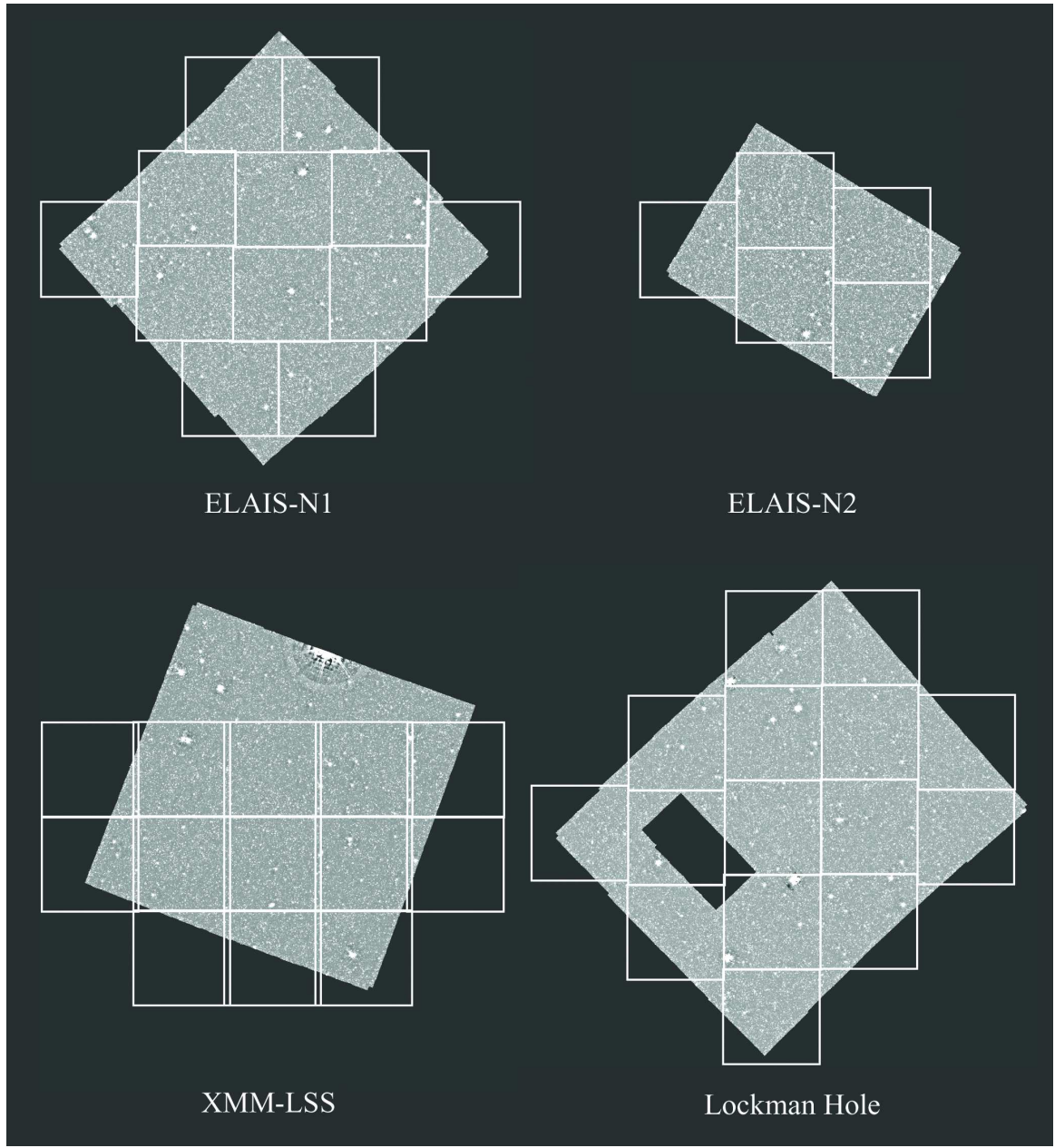}
\caption{\footnotesize The 3.6$\micron$ mosaics for the four SWIRE fields
  observable from the northern hemisphere.  The location of the SpARCS
  z$^{\prime}$-band CFHT/MegaCam pointings are overplotted as white
  boxes.  Each MegaCam pointing covers $\sim$ 1 deg$^2$.  The observations of the XMM-LSS field were obtained as part of the
  CFHTLS-wide survey.  Excluding areas masked by bright stars and
  missed by MegaCam chip gaps there are 28.3 deg$^2$ with both
  z$^{\prime}$ and 3.6$\micron$ observations in the northern fields that can
  be used for cluster finding.}
\end{figure*}
\section{Cluster Selection}
\indent
Clusters are found in the data using the cluster red-sequence
algorithm developed by Gladders \& Yee (2000; 2005).  Muzzin et
al. (2008) used a slightly modified version of the algorithm to
detect clusters at 0 $< z <$ 1.3 in the
FLS using an R - 3.6$\micron$ color.  We use the  Muzzin et
al. (2008) code for the SpARCS data.  The change from a R -
3.6$\micron$ color to a $z^{\prime}$ - 3.6$\micron$ is optimum for
targeting clusters at $z >$ 1, where the z$^{\prime}$ - 3.6$\micron$  
color spans the 4000\AA~break.  Other than the change of using a
different optical band, the
SpARCS algorithm is identical to that presented in  Muzzin et
al. (2008) and we refer to that paper for further details of the
cluster finding technique.
\newline\indent
After the first semester of $z^{\prime}$ observations were complete
there were $\sim$ 14 deg$^2$ of data with both z$^{\prime}$ and 3.6$\micron$
data.  From this area we selected two rich cluster candidates, both
from the ELAIS-N2 field with red-sequence photometric redshifts\footnote{Based on a
$z_{f}$ = 2.8 single-burst Bruzual \& Charlot (2003) model} of $z >$
1.2 for spectroscopic followup.
These two cluster candidates, SpARCS J163435+402151 (R.A.: 16:34:35.0,
Decl:+40:21:51.0), and SpARCS J163852+403843 (R.A.:16:38:52.0, Decl:+40:38:43.0), have
richnesses, parameterized by B$_{gc,R}$, of 1053 $\pm$ 278 Mpc$^{1.8}$
and 988 $\pm$ 270 Mpc$^{1.8}$, respectively.  For a discussion of B$_{gc}$ and B$_{gc,R}$ as cluster
richness estimates see Yee \& Lopez-Cruz (1999), and Gladders \& Yee (2005).
Based on the empirical calibration of B$_{gc}$ vs. M$_{200}$ determined
by Muzzin et al. (2007) in the K-band for the CNOC1 clusters at $z
\sim$ 0.3, this implies M$_{200}$ = 5.7 x 10$^{14}$ M$_{\odot}$ and
5.1 x 10$^{14}$ M$_{\odot}$ for SpARCS J163435+402151 and SpARCS
J163852+403843, respectively.  Although Muzzin et al. (2007) found
there is a fairly large scatter in the B$_{gc}$ vs. M$_{200}$
relation, the high richnesses  imply that these candidates are 
likely to be massive, high-redshift systems.
\begin{figure}
\includegraphics[scale = 0.44]{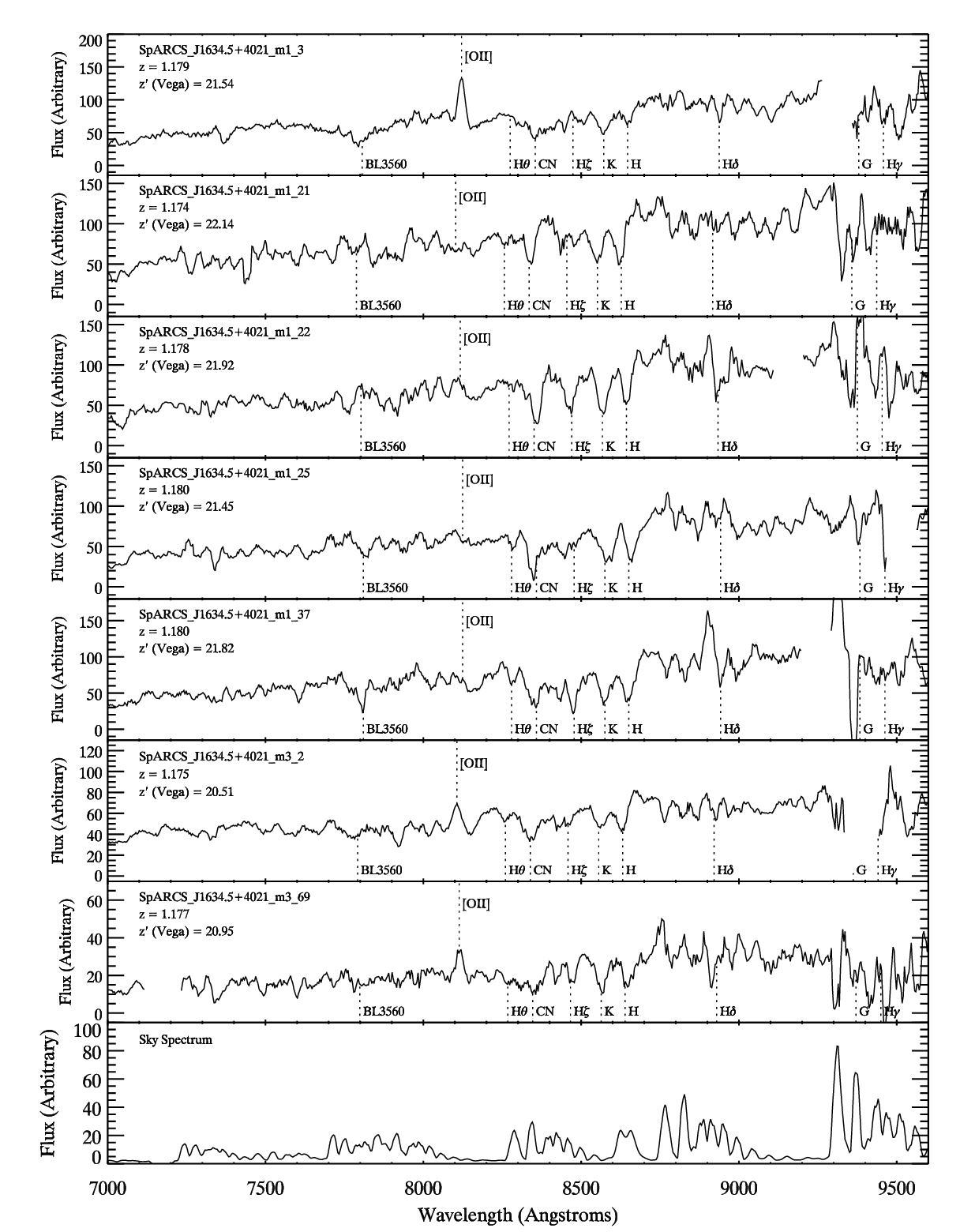}
\caption{\footnotesize Spectra for a subsample of seven galaxies in
  the cluster SpARCS J163435+402151.  The spectra have been smoothed with a
  7-pixel boxcar so the sampling matches the instrumental resolution.}
\end{figure}
\begin{figure}
\includegraphics[scale = 0.44]{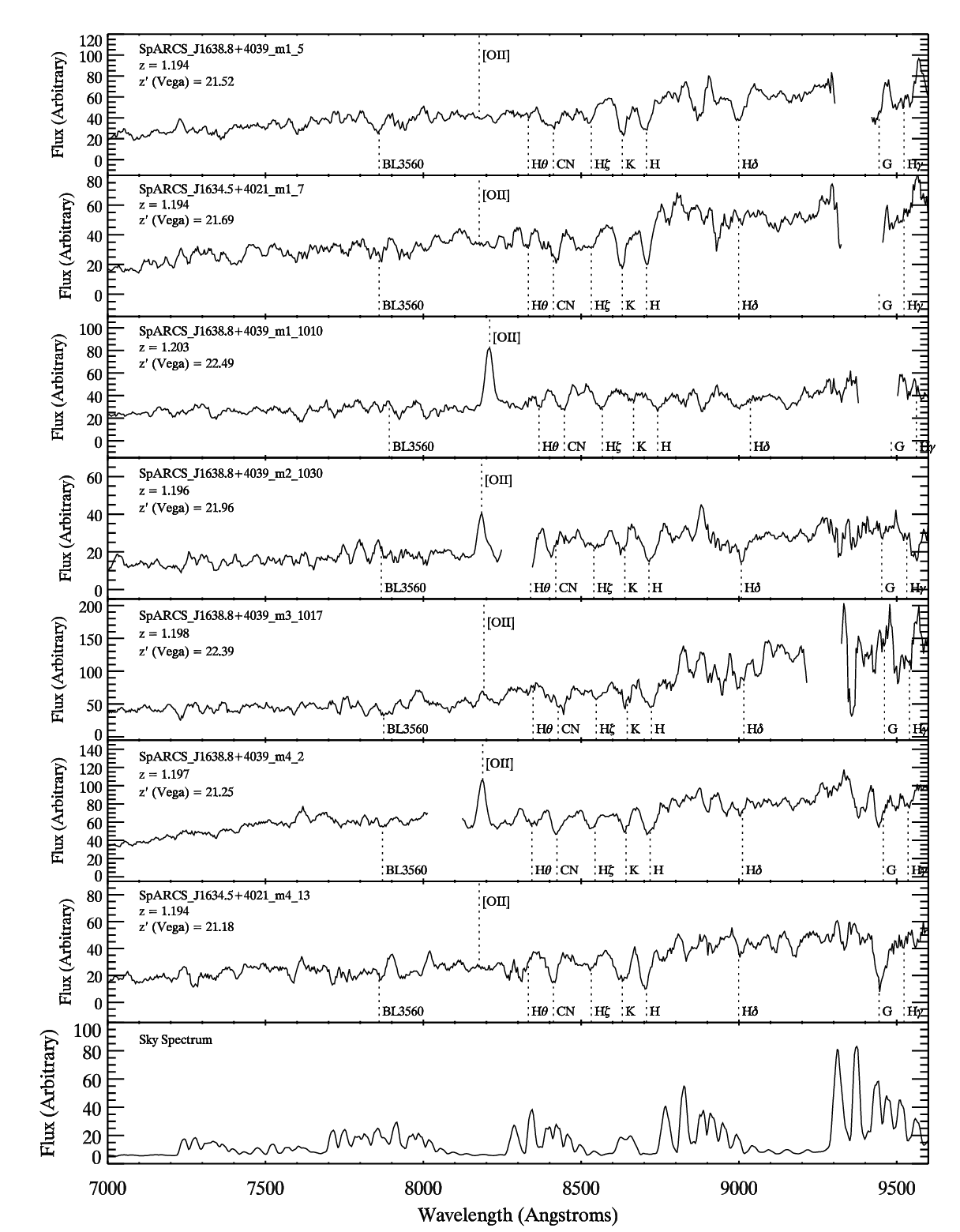}
\caption{\footnotesize Same as Figure 2 but for galaxies in 
  the cluster SpARCS J163852+403843.  }
\end{figure}
\section{Spectroscopic Data}
Multislit nod-and-shuffle (N\&S) spectroscopy of galaxies in SpARCS J163435+402151, and SpARCS J163852+403843 were obtained
 using GMOS-N on Gemini as part of the program GN-2007A-Q-17.  We used the R150 grating blazed at 7170A with 1$''$
width slits. This provided a resolving power of R = 631 which
 corresponds to a resolution of $\sim$ 11\AA, or $\sim$ 250 km s$^{-1}$ at the
estimated redshift of the clusters.  For all observations we used
 3$''$ long microslits, corresponding to roughly four seeing-disks, 
  allowing a two seeing-disk spacing between the nod
 positions.   We observed three masks for SpARCS
 J163435+402151 and four masks for SpARCS J163852+403843.  One mask
 for each of the clusters was observed in ``micro-shuffle'' mode,
 but the majority were observed in ``band-shuffle'' mode.  All masks were
 observed using the RG615 filter which blocks light blueward of
 6150\AA~so that multiple tiers of slits could be used.
\newline\indent
Unlike micro-shuffle where the shuffled charge is stored directly adjacent to
 the slit location, band-shuffle shuffles the charge to the top and
 bottom third of the chips for storage.  While  technically  it is the least-efficient N\&S mode in terms of usable
 area for observations (only the central 1.7$'$ of the total 5$'$
 FOV can be used) it is extremely efficient for observations of
 high-redshift clusters because it allows the microslits to be packed
 directly beside each other in the cluster core with no requirement for additional space for
 storing the shuffled charge.  In band-shuffle mode we were typically able to
 locate between 20-26 slits, including three
   alignment stars, per mask in the central 1.7$'$ around the
 cluster.  At $z \sim$ 1.2 the 1.7$'$
 FOV corresponds to a diameter of 850 kpc, roughly the projected size of a massive cluster.
\newline\indent
Slits were placed on galaxies with priorities in the following order:
Priority 1, galaxies with colors $\pm$ 0.6 mag from the red-sequence and 3.6$\micron$
$<$ 16.9.  Priority 2, galaxies with colors $\pm$ 0.6 mag from the red-sequence, 3.6$\micron$ $>$ 16.9 and
z$^{\prime}$ $<$ 23.5.  Priority 3, galaxies with colors $>$ 0.6 bluer than the red-sequence, but $<$ 1.0 mag
bluer and 3.6$\micron$ $>$ 16.9 and z$^{\prime}$ $<$ 23.5.  Priority
4, same as priority 3 but for galaxies with colors bluer than the
red-sequence by 1.0-1.4 mag.  Priority 5: all galaxies with 23.5 $<$
z$^{\prime}$ $<$ 24.5.  
Roughly speaking, Priorities 1 through 4 can be described as bright
red-sequence, faint red-sequence, blue cloud, and extreme blue cloud
galaxies, respectively.   
\newline\indent
For each mask we obtained a total of 3 hrs of integration time by combining six exposures with 30 mins of 
integration time.  The six frames were obtained using 15 nod cycles of 60s integration time per
cycle.  Each of the 6 exposures was offset by a few arcseconds using
the on-chip dithering option.
\subsection{Data Reduction}
Data were reduced using the GMOS IRAF package.  We subtracted a bias
and N\&S dark from each frame.  The N\&S darks are taken using the
same exposure times and using the same charge shuffling routine as the science observations, but
with the shutter closed.  Regions with poor charge transfer efficiency
cause electrons to become trapped during the repeated charge shuffling
used in the observations.  Such charge traps can be identified and
corrected using dark frames taken with the same N\&S settings.  
Images were registered using bright sky lines and sky subtracted using the complementary storage area using the
``gnsskysub'' task.  Final mosaics are made by coadding the
sky subtracted images.
\newline\indent
One dimensional spectra were extracted using the iGDDS software
(Abraham et al. 2004).  Wavelength calibration for each extracted
spectrum was performed using bright sky lines from the unsubtracted
image, also with the iGDDS software.  Wavelength solutions typically
have an rms $<$ 0.5\AA.  We determined a relative flux calibration
curve using a long slit observation of the standard star EG131.
\newline\indent
Redshifts were determined interactively for each spectrum by comparing with the
templates available in iGDDS. Most of the redshifts were identified
using the early-, intermediate-, and late-type composite spectra
from the Gemini Deep Deep Survey (Abraham et al. 2004).
The final redshifts was determined using the average redshift from
all absorption and emission lines that were detected.  The vast
majority of redshifts were determined by identifying the  [OII] 3727\AA-doublet
emission line
(which is not resolved at our resolution), or the Calcium II H+K
absorption lines.  Many of the spectra also show the Balmer series
lines.   We list the spectroscopic members of SpARCS
J163435+402151 and SpARCS J163852+403843 in Tables 1 and 2, and the
spectroscopically confirmed foreground/background galaxies in Tables 3
and 4.  We also plot examples of some cluster galaxy spectra in
Figures 2 and 3.  R, z$^{\prime}$, and 3.6$\micron$ color composites of the
two clusters are shown in
Figures 4 and 5.  The white squares denote the spectroscopically
confirmed cluster members and the green squares denote the
spectroscopically confirmed foreground/background galaxies.
\begin{center}
\begin{deluxetable}{lcccc}
\tabletypesize{\footnotesize}
\scriptsize
\tablecaption{Spectroscopic Cluster Members in SpARCS J163435+402151}
\tablehead{\colhead{ID} & \colhead{R.A.} & \colhead{Decl.} &
  \colhead{z$^{\prime}$} & \colhead{$z_{spec}$}\\
 \colhead{} & \colhead{J2000 (Deg.)} & \colhead{J2000 (Deg.)} &
  \colhead{Mag Vega} & \colhead{}
}
\startdata
\multicolumn{5}{c}{Mask 1}\\
\hline
     3	& 248.6589 &  40.35303	& 21.54 &  1.179 \\
     5	& 248.6497 &  40.36148	& 21.58 &  1.181 \\
     6	& 248.6467 &  40.36418	& 20.95 &  1.166 \\
    21	& 248.6513 &  40.35238	& 22.14 &  1.174 \\
    22	& 248.6605 &  40.35474	& 21.91 &  1.178 \\
    25	& 248.6510 &  40.35855	& 21.44 &  1.180 \\
    28	& 248.6605 &  40.36132	& 22.19 &  1.182 \\
    34	& 248.6527 &  40.36477	& 22.78 &  1.182 \\
    37	& 248.6496 &  40.36696	& 21.82 &  1.180 \\
    41	& 248.6459 &  40.36841	& 21.97 &  1.185 \\
    45	& 248.6440 &  40.37413	& 22.14 &  1.187 \\
  3027	& 248.6546 &  40.35461	& 21.59 &  1.181 \\
\hline
\multicolumn{5}{c}{Mask 2}\\
\hline
    20  & 248.5991 &  40.34870	& 23.10 &  1.170\\
    27	& 248.6548 &  40.35104	& 22.23 &  1.164\\
    36	& 248.6171 &  40.36649	& 22.36 &  1.184\\
    57	& 248.6700 &  40.39159	& 22.80 &  1.176\\
\hline
\multicolumn{5}{c}{Mask 3}\\
\hline
     2 & 248.6584 &  40.34934 & 20.50 &  1.175\\
    38 & 248.6474 &  40.36152 & 21.67 &  1.178\\
    69 & 248.6669 &  40.39132 & 21.89 &  1.178
\enddata
\end{deluxetable}
\begin{deluxetable}{lcccc}
\tabletypesize{\footnotesize}
\scriptsize
\tablecaption{Spectroscopic Cluster Members in SpARCS J163852+403843}
\tablehead{\colhead{ID} & \colhead{R.A.} & \colhead{Decl.} &
  \colhead{z$^{\prime}$} & \colhead{$z_{spec}$}\\
 \colhead{} & \colhead{J2000 (Deg.)} & \colhead{J2000 (Deg.)} &
  \colhead{Mag Vega} & \colhead{}
}
\startdata
\multicolumn{5}{c}{Mask 1}\\
\hline
     5	& 249.7132 &  40.63952 	& 21.51 &  1.194\\
     7	& 249.7144 &  40.64159	& 21.68 &  1.194\\
     8	& 249.7152 &  40.64527	& 21.41 &  1.195\\
  1007	& 249.6974 &  40.63383	& 22.50 &  1.199\\
  1010	& 249.7151 &  40.63804	& 22.48 &  1.202\\
  1016	& 249.7006 &  40.64276	& 22.20 &  1.200\\
  1025	& 249.7137 &  40.64952	& 22.46 &  1.200\\
  1028	& 249.7104 &  40.65532	& 21.96 &  1.190\\
  1031	& 249.7029 &  40.65968	& 22.85 &  1.176\\
  3026	& 249.6668 &  40.64420	& 22.73 &  1.195\\
\hline
\multicolumn{5}{c}{Mask 2}\\
\hline
  1018	& 249.7032 &  40.64533	& 22.67 &  1.186\\
  1020	& 249.7099 &  40.64631	& 22.39 &  1.202\\
  1024	& 249.7135 &  40.64828	& 22.24 &  1.198\\
  1026	& 249.7126 &  40.65311	& 22.26 &  1.200\\
  1030	& 249.7581 &  40.63628	& 21.95 &  1.195\\
  2016	& 249.6610 &  40.65976	& 22.22 &  1.197\\
\hline
\multicolumn{5}{c}{Mask 3}\\
\hline
  1017	& 249.7136 &  40.64511	& 22.38 &  1.198\\
  1019	& 249.7163 &  40.64602	& 22.29 &  1.198\\
  2009	& 249.7236 &  40.63171	& 22.10 &  1.188\\
  4029	& 249.7619 &  40.65497	& 23.67 &  1.195\\
\hline
\multicolumn{5}{c}{Mask 4}\\
\hline
     2	&  249.6947 &  40.61541	& 21.24 &  1.196\\
     3	&  249.6986 &  40.62432	& 21.73 &  1.200\\
    10	&  249.7314 &  40.66114	& 21.25 &  1.192\\
    11	&  249.7378 &  40.66462	& 21.72 &  1.194\\
    13	&  249.6680 &  40.67089	& 21.17 &  1.194\\
  2014	&  249.7532 &  40.64920	& 22.58 &  1.196\\
  2030	&  249.7049 &  40.68847	& 21.87 &  1.175\\
  1012	&  249.6992 &  40.63979	& 23.06 &  1.196\\
  1021	&  249.7217 &  40.64632	& 21.90 &  1.195\\
  3019	&  249.7189 &  40.63320	& 23.17 &  1.219\\
  3046	&  249.7509 &  40.68401	& 22.05 &  1.172
\enddata
\end{deluxetable}
\begin{deluxetable}{lcccc}
\tabletypesize{\footnotesize}
\scriptsize
\tablecaption{Spectroscopic Foreground/Background Galaxies in Field of
  SpARCS J163435+402151}
\tablehead{\colhead{ID} & \colhead{R.A.} & \colhead{Decl.} &
  \colhead{z$^{\prime}$} & \colhead{$z_{spec}$}\\
 \colhead{} & \colhead{J2000 (Deg.)} & \colhead{J2000 (Deg.)} &
  \colhead{Mag Vega} & \colhead{}
}
\startdata
\multicolumn{5}{c}{Mask 1}\\
\hline
     4	& 248.6129 &  40.35610	& 20.89 &  1.108 \\
    26	& 248.6600 &  40.35898	& 21.38 &  1.004 \\
    31	& 248.6042 &  40.36213	& 21.66 &  1.105 \\
    45	& 248.6823 &  40.38687	& 23.64 &  0.925 \\
\hline
\multicolumn{5}{c}{Mask 2}\\
\hline
    23 & 248.60280 &  40.33815 & 23.79 & 1.337\\
    24 & 248.60710 &  40.33762 & 24.05 & 1.255\\
\hline
\multicolumn{5}{c}{Mask 3}\\
\hline
     1	& 248.67280 &  40.33252	& 21.16 &  1.348\\
    15	& 248.65750 &  40.33183	& 21.32 &  0.780\\
    65	& 248.67920 &  40.38581	& 23.45 &  1.108\\
    67	& 248.63340 &  40.38773	& 23.05 &  0.811
\enddata
\end{deluxetable}
\begin{deluxetable}{lcccc}
\tabletypesize{\footnotesize}
\scriptsize
\tablecaption{Spectroscopic Foreground/Background Galaxies in Field of SpARCS J163852+403843}
\tablehead{\colhead{ID} & \colhead{R.A.} & \colhead{Decl.} &
  \colhead{z$^{\prime}$} & \colhead{$z_{spec}$}\\
 \colhead{} & \colhead{J2000 (Deg.)} & \colhead{J2000 (Deg.)} &
  \colhead{Mag Vega} & \colhead{}
}
\startdata
\multicolumn{5}{c}{Mask 1}\\
\hline
  3021 &  249.7659 &  40.63502 & 22.14 &  0.670\\
\hline
\multicolumn{5}{c}{Mask 2}\\
\hline
  3024 & 249.7503 &  40.64119 & 23.61 &  0.875\\
  4020 & 249.7305 &  40.63887 & 23.98 &  1.391\\
  4030 & 249.7666 &  40.65555 & 23.56 &  0.776\\
\hline
\multicolumn{5}{c}{Mask 3}\\
\hline
  3034 & 249.6953 &  40.65273 &	21.96 &  1.393\\
  3036 & 249.6603 &  40.65813 &	23.47 &  1.386\\
\hline
\multicolumn{5}{c}{Mask 4}\\
\hline
     1 & 249.7320 &  40.60904 &	22.10 &  0.963\\
    12 & 249.6992 &  40.66802 &	20.96 &  0.784\\
  2003 & 249.6765 &  40.60412 &	22.44 &  1.017\\
  2024 & 249.7496 &  40.67251 &	23.50 &  0.848\\
  2026 & 249.6744 &  40.67759 &	21.46 &  0.771\\
  3004 & 249.7519 &  40.60677 & 21.44 &  0.768\\
  3041 & 249.7611 &  40.67033 &	21.17 &  0.784
\enddata
\end{deluxetable}
\end{center}
\begin{figure*}
\plotone{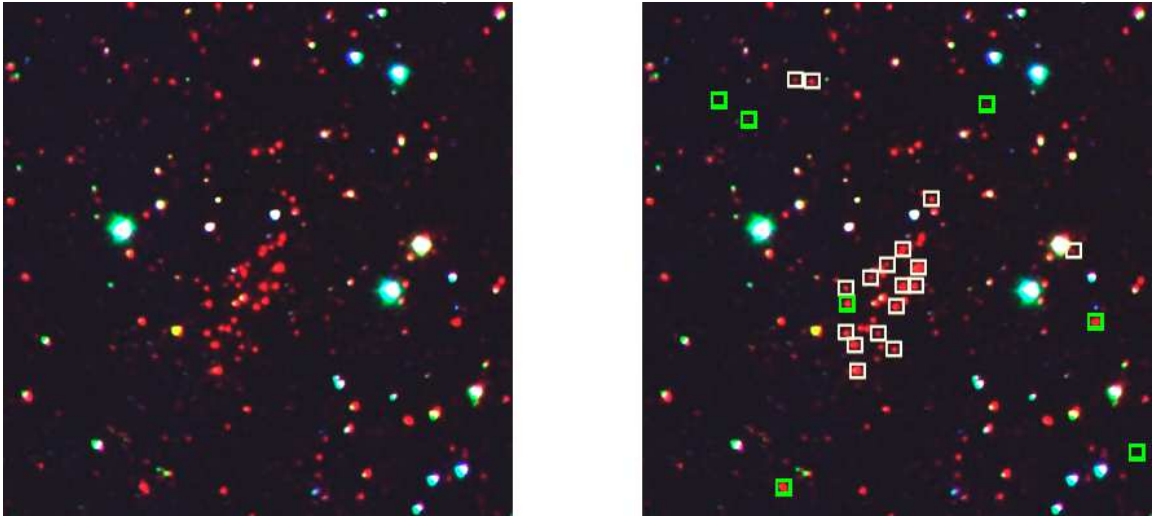}
\caption{\footnotesize Left: Rz$^{\prime}$3.6$\micron$ color composite of
  the cluster SpARCS J163435+402151 at $z =$ 1.1798.  The R and z$^{\prime}$ images
  have been convolved to match the 3.6$\micron$ PSF.  The FOV of the
  image is $\sim$ 3.5$'$ across.  Right: Same as
  left panel but with 
  spectroscopically confirmed cluster members marked as white squares
  and spectroscopically confirmed foreground/background galaxies marked
  as green squares.}
\end{figure*}
\begin{figure*}
\plotone{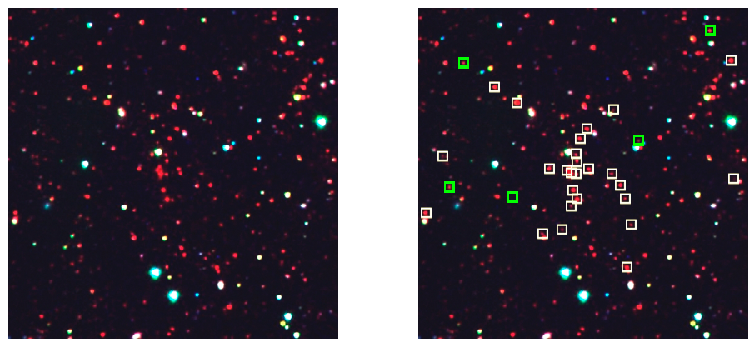}
\caption{\footnotesize As Figure 4, but for the cluster SpARCS
  J163852+403843 at $z =$ 1.1963.  The FOV of the images is $\sim$
  4.5$'$ across.}
\end{figure*}
\section{Cluster Velocity Dispersions}
For both clusters there are a sufficient number of redshifts to determine a velocity
dispersion ($\sigma_{v}$) and therefore a dynamical mass.  Our $\sigma_{v}$'s are
determined using the method detailed in Blindert (2006).  Briefly, we
make a rejection of near-field non-cluster members using a modified version
of the Fadda et al. (1996) shifting-gap procedure.  This method uses
both the position and velocity of galaxies to reject interlopers.  In
Figure 6 we plot the relative velocities of the cluster galaxies as a
function of projected radius. Two galaxies in SpARCS J163435+402151 are rejected as near-field
interlopers and three are rejected in SpARCS J163852+403843.
The rejected galaxies are plotted as crosses in Figure 6 and are not used in
computing the mean redshift of the cluster or $\sigma_{v}$.  Once outliers are rejected the redshift of the clusters
is determined using the remaining galaxies.  The spectroscopic redshift of the
clusters is 1.1798 and 1.1963 for SpARCS J163435+40215 and SpARCS
J163852+403843, respectively.
\newline\indent
The $\sigma_{v}$'s are determined using the ``robust'' estimator
suggested by Beers et al. (1990) and Girardi et al. (1993).  The
robust estimator is simply the biweight estimator for systems with $>$
15 members, and the gapper estimator for systems with $<$ 15 members.
As discussed in those papers and Blindert (2006) these estimators are
more robust than standard deviations as they are less sensitive to
outliers, which may still persist even after the initial shifting-gap rejection.
Using the ``robust'' estimator, SpARCS J163435+402151 and SpARCS J163852+403843 have $\sigma_{v}$ = 490 $\pm$ 140 km s$^{-1}$
and 650 $\pm$ 160 km s$^{-1}$, respectively, where the errors have
been determined using Jackknife resampling of the data.  
\newline\indent
We estimate the dynamical
mass using M$_{200}$, the mass contained within r$_{200}$, the radius at which the
mean interior density is 200 times the critical density ($\rho_{c}$).  We use the
equation, 
\begin{equation}
M_{200} = \frac{4}{3}\pi r_{200}^3 \cdot 200\rho_{c},    
\end{equation}
with the dynamical estimate of r$_{200}$ from Carlberg et al. (1997),
\begin{equation}
r_{200} = \frac{\sqrt{3}\sigma}{10H(z)},
\end{equation}
where H($z$) is the Hubble constant at the redshift of the cluster.
From these relation we derive r$_{200}$ = 0.62 $\pm$ 0.18 Mpc, and 0.82 $\pm$
0.20 Mpc for SpARCS J163435+402151 and SpARCS J163852+403843,
respectively.  From Equation 2, these imply M$_{200}$ = (1.0 $\pm$
0.9) x 10$^{14}$ M$_{\odot}$ and (2.4 $\pm$
1.8) x 10$^{14}$ M$_{\odot}$ for SpARCS J163435+402151 and SpARCS J163852+403843,
respectively.
\begin{figure*}
\plottwo{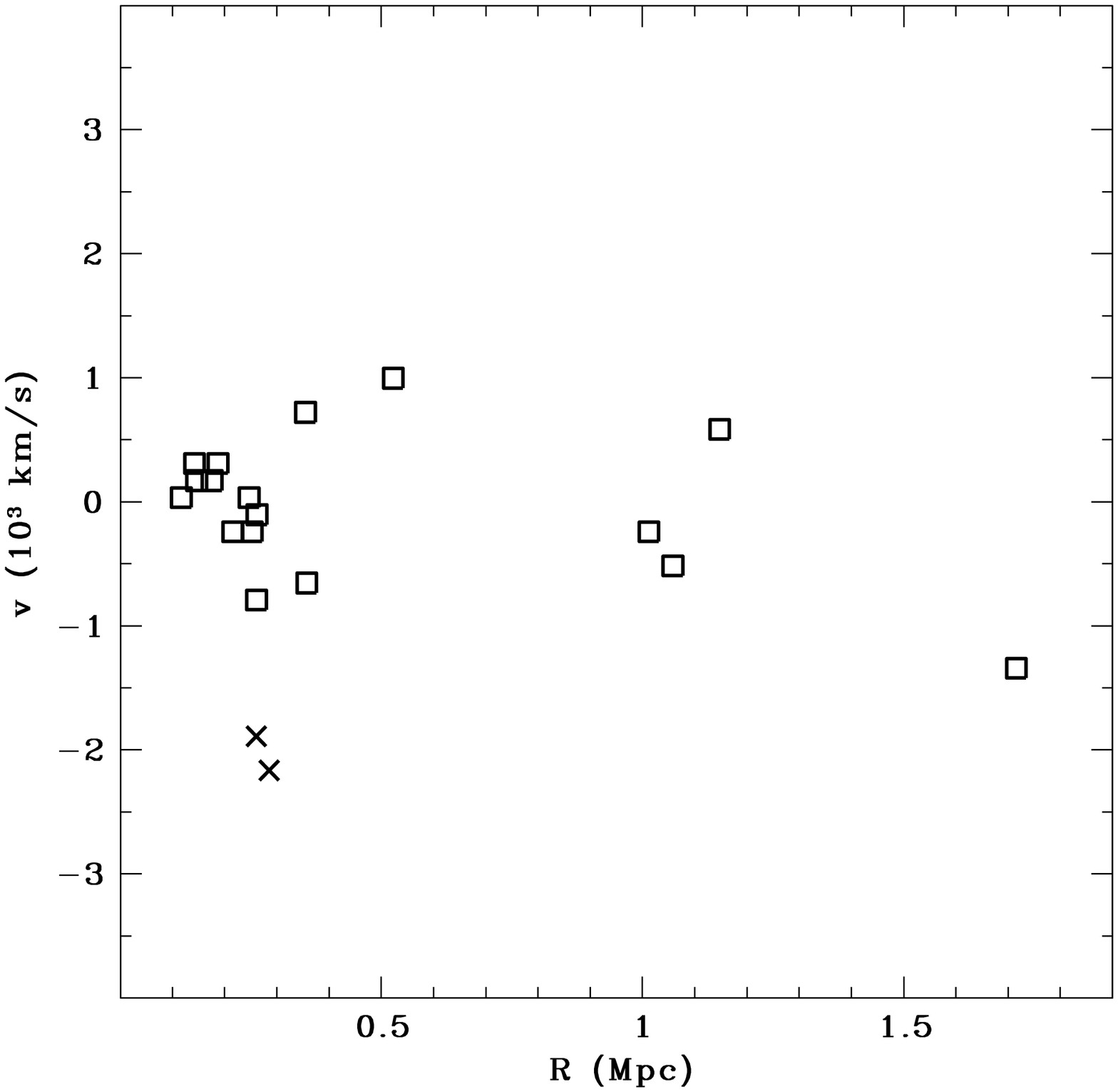}{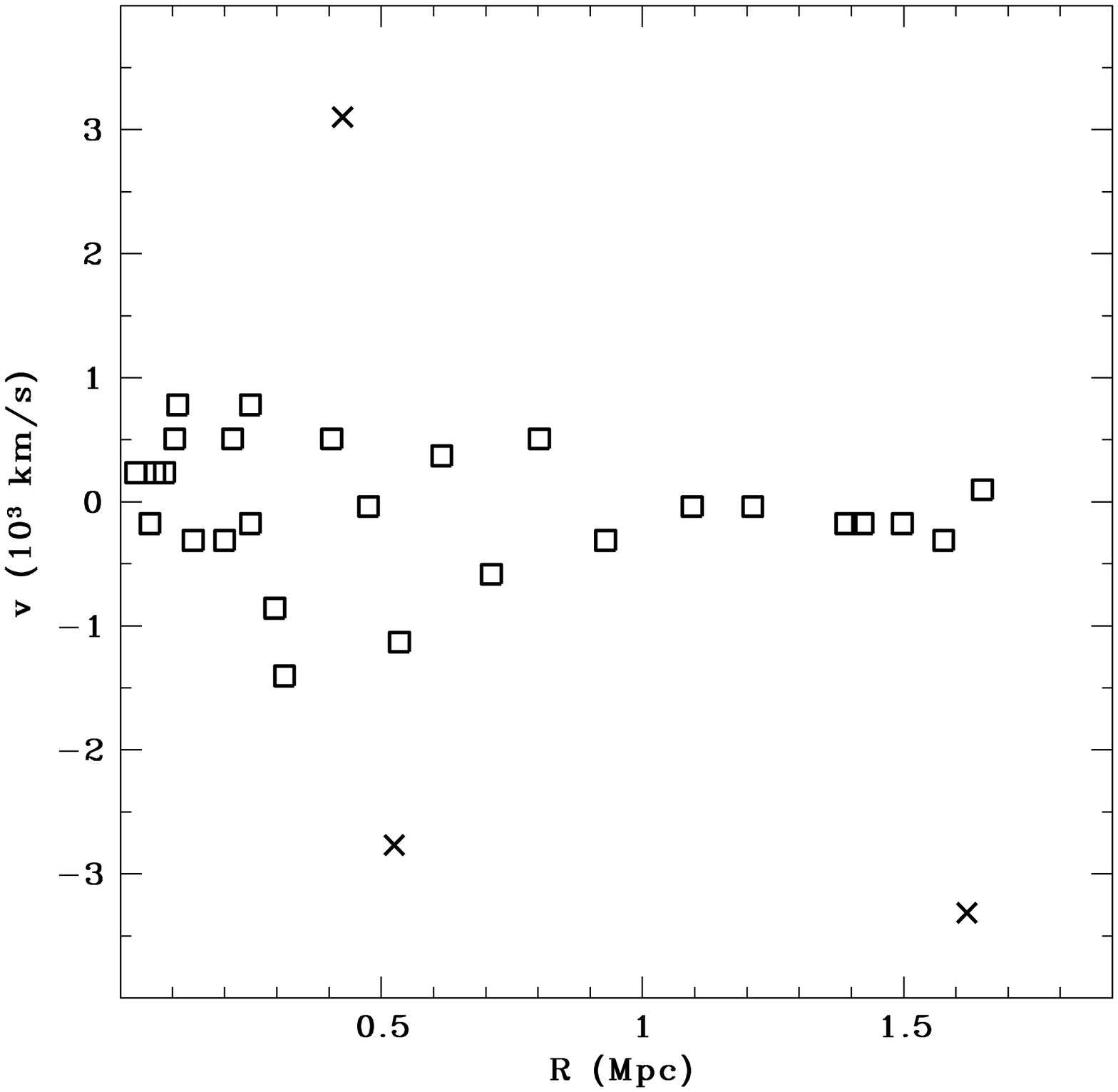}
\caption{\footnotesize Left Panel: Galaxy velocities relative to the
  cluster mean velocity as a function of
  radius for SpARCS J163435+402151.  Right Panel: Same as left panel but
  for SpARCS J163852+403843.
  Galaxies marked with an ``x'' are more likely to be near-field
  objects than members of the cluster and are not used in the
  computation of the velocity dispersion.}
\end{figure*}
\begin{figure*}
\plottwo{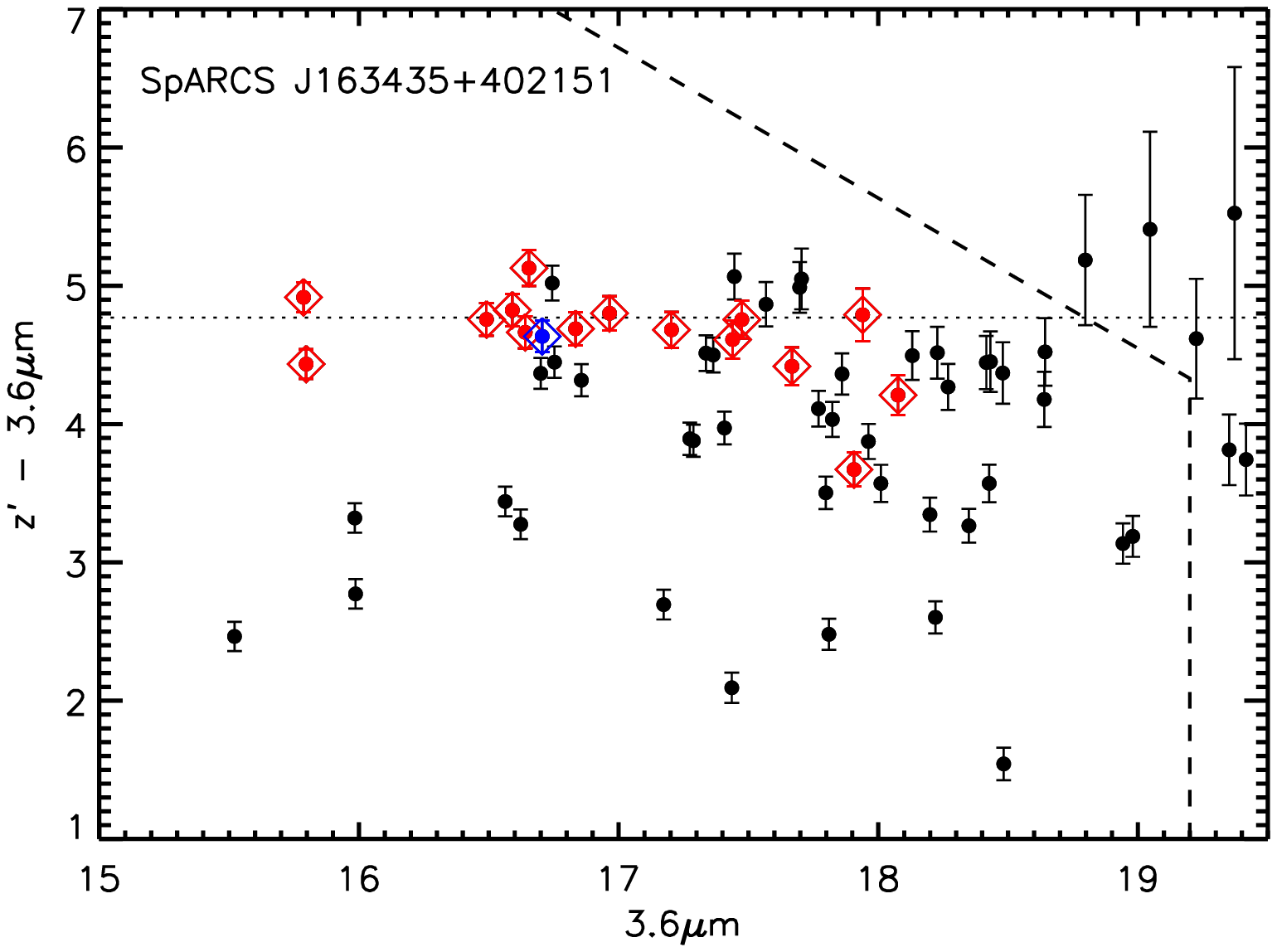}{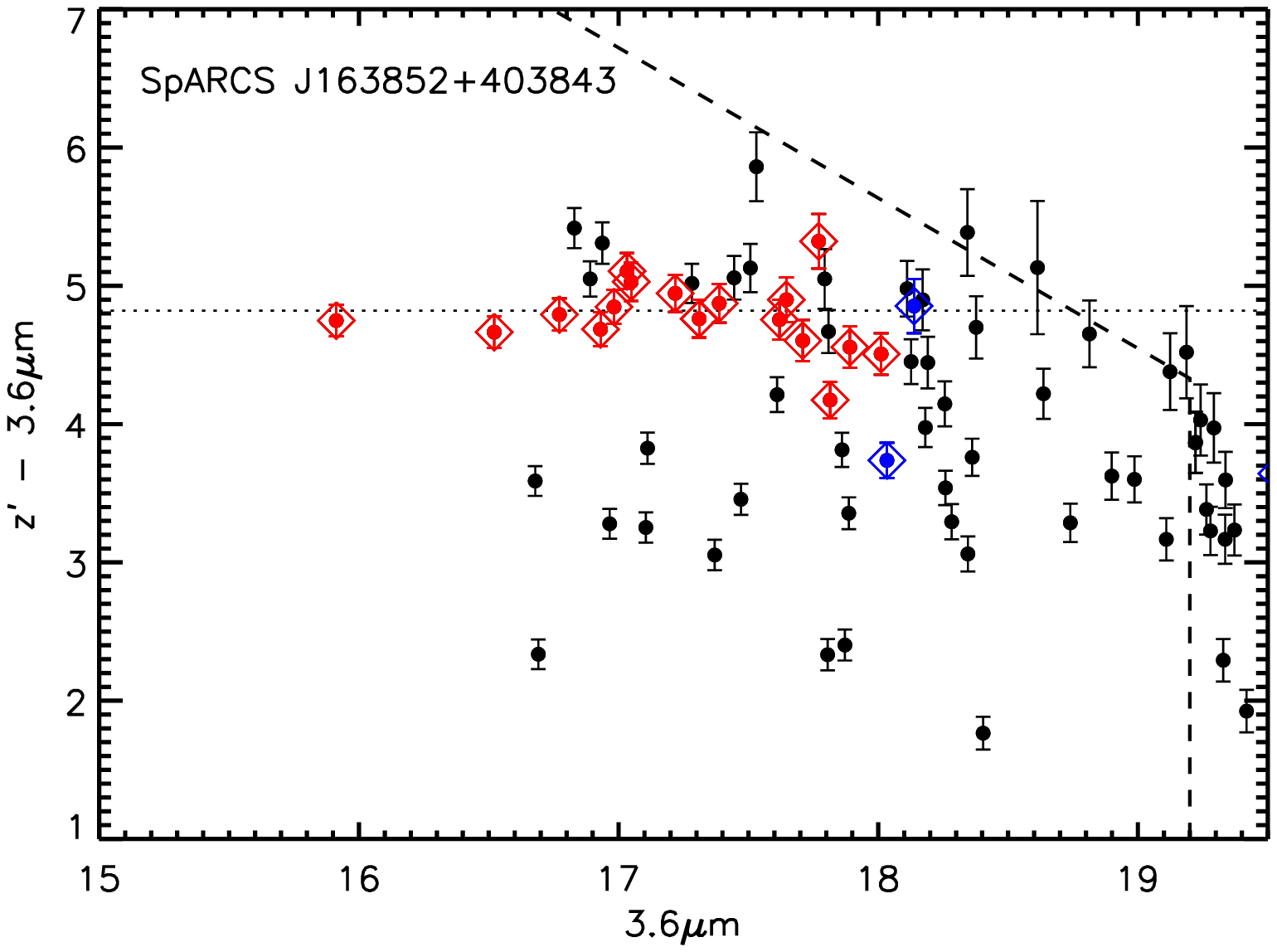}
\caption{\footnotesize Left Panel: z$^{\prime}$ - 3.6$\micron$
  vs. 3.6$\micron$ color magnitude diagram for galaxies at R $<$ 550
  kpc in the field of the cluster SpARCS J163435+402151.
  Spectroscopically confirmed cluster members and
  foreground/background galaxies are plotted as red and blue diamonds,
  respectively.  The dotted line is the best fit line 
  to the confirmed cluster members with the slope fixed at zero.  Right Panel: Same as left panel but
  for SpARCS J163852+403843.}
\end{figure*}
\section{Discussion}
\subsection{Red-sequence Photometric Redshifts}
\indent
 In Figure 7 we plot the z$^{\prime}$ - 3.6$\micron$ vs. 3.6$\micron$ color magnitude relation
for galaxies at projected radii (R) $<$ 550 kpc in the fields of both clusters.  Spectroscopically
confirmed members and confirmed foreground/background galaxies are plotted as red and blue diamonds,
respectively.  The dotted line in both panels is the best fit 
 line with the slope fixed at zero, to the spectroscopically confirmed members.  These lines
 indicate that the red-sequence galaxies have z$^{\prime}$ -
 3.6$\micron$ colors of 4.77 and 4.82 for  SpARCS J163435+402151 and
 SpARCS J163852+403843, respectively.  Using a solar metallicity, Bruzual \& Charlot
 (2003) simple stellar population (SSP) with a z$_{f}$ = 4.0 these
 colors imply photometric redshifts of 1.19 and 1.20, in excellent
 agreement with the spectroscopic redshifts.  At $z \sim$ 1.2, the
 red-sequence photometric redshifts do not depend strongly on the
 chosen $z_{f}$.  If we instead use a z$_{f}$ = 2.8 SSP, the
 red-sequence color would predict redshifts of 1.21 and 1.24, and for
 a $z_{f}$ = 10.0 SSP it would predict redshifts of 1.13 and 1.15.
 However, the color differences between all these models at fixed redshift are small
 ($<$ 0.1 mag), and so it is not possible to distinguish between different
 formation epochs without more data.
 Still, the close agreement between the red-sequence
 photometric redshift derived using a reasonable $z_{f}$ and the spectroscopic redshift is 
 encouraging for the use of red-sequence photometric
 redshifts for clusters at $z >$ 1.
\subsection{Mass vs. Richness}
\indent
Both clusters have lower masses than predicted by their richness by factors of $\sim$ 6 and 2
for SpARCS J163435+402151 and SpARCS J163852+403843, respectively,
although due to the large error bars the differences are only
significant at $\sim$ 1 and 2$\sigma$, respectively.
Whether this represents a redshift evolution in the B$_{gc}$-M$_{200}$
scaling relation, or is simply a richness-selected Eddington bias\footnote{We followed
up two of the richest clusters in our early dataset.  The cluster mass
function is steep at high redshift and low mass systems greatly
outnumber high mass systems.  Due to scatter in the mass-richness
relation lower mass systems  with
abnormally high richnesses may be more common than truly massive systems. }
is impossible to determine using only two clusters.  Both Gilbank et
al. (2007) and Andreon et al. (2007) found that for a small sample of clusters at $z
\sim$ 1 the cluster richnesses were still consistent with their
velocity dispersions based on relations calibrated at lower redshift,
although both parameters have large uncertainties in their measurements.  More clusters with
well-determined $\sigma_{v}$ and B$_{gc}$ will be needed to test if the
cluster scaling relations at $z >$ 1 are similar to those at lower redshift.
\section{Summary}
\indent
We have presented a brief summary of observations for the northern
component of the SpARCS survey.  Using Gemini N\&S spectroscopy we
confirmed two rich cluster candidates at $z \sim$ 1.2 selected from
early survey data.  We find that the photometric redshifts from the
color of the cluster red-sequence agree extremely well with the
spectroscopic redshifts.  Both clusters have a smaller M$_{200}$
than would be expected from their richness if we use the B$_{gc}$ -
M$_{200}$ scaling relation calibrated at $z \sim$ 0.3.  Whether this
represents a true evolution in the cluster scaling relations at $z >$
1.2 or is simply a selection bias will require well-determined
M$_{200}$ for a larger sample of clusters.
\newline\indent
Overall, the confirmation of both SpARCS J163435+402151 and SpARCS J163852+403843
as {\it bona fide} massive clusters at $z >$ 1 provide strong evidence
that the  red-sequence technique  is  an  effective and
efficient method for detecting clusters  at $z >$
1 (see also Wilson et al. 2008 who present a confirmed $z =$ 1.34
cluster from the southern component of the SpARCS survey).  The
complete SpARCS catalogue contains $hundreds$ of cluster candidates at
$z >$ 1 and promises to be one of the premier data sets for the study
of cluster galaxy evolution at $z >$ 1.


\acknowledgements

Based on observations obtained with MegaPrime/MegaCam, a joint project
of CFHT and CEA/DAPNIA, at the Canada-France-Hawaii Telescope (CFHT)
which is operated by the National Research Council (NRC) of Canada,
the Institut National des Sciences de l'Univers of the Centre National
de la Recherche Scientifique (CNRS) of France, and the University of
Hawaii. This work is based in part on data products produced at
TERAPIX and the Canadian Astronomy Data Centre as part of the
Canada-France-Hawaii Telescope Legacy Survey, a collaborative project
of NRC and CNRS.

Based on observations obtained at the Gemini Observatory, which is operated by the
Association of Universities for Research in Astronomy, Inc., under a cooperative agreement
with the NSF on behalf of the Gemini partnership: the National Science Foundation (United
States), the Science and Technology Facilities Council (United Kingdom), the
National Research Council (Canada), CONICYT (Chile), the Australian Research Council
(Australia), Ministerio da Ciencia e Tecnologia (Brazil) and SECYT (Argentina)

This work is based in part on observations made with the Spitzer Space
Telescope, which is operated by the Jet Propulsion Laboratory, California
Institute of Technology under a contract with NASA.

\end{document}